\documentclass[reprint,aip,apl]{revtex4-2}
\usepackage{graphicx}% Include figure files
\usepackage{dcolumn}% Align table columns on decimal point
\usepackage{bm}% bold math
\usepackage{amsmath}
\usepackage{amssymb}
\usepackage[utf8]{inputenc}
\usepackage[T1]{fontenc}
\usepackage{mathptmx}
\usepackage{etoolbox}
\usepackage{siunitx}

\DeclareSIUnit\angstrom{\text{\AA}}

\makeatletter
\def\@email#1#2{%
 \endgroup
 \patchcmd{\titleblock@produce}
  {\frontmatter@RRAPformat}
  {\frontmatter@RRAPformat{\produce@RRAP{*#1\href{mailto:#2}{#2}}}\frontmatter@RRAPformat}
  {}{}
}%
\makeatother
\begin{document}

\preprint{AIP/123-QED}

\title[Multilayer Laue Lenses for Enhanced Spatial Resolution in Dark-Field X-ray Microscopy]{Multilayer Laue Lenses for Enhanced Spatial Resolution in Dark-Field X-ray Microscopy}

\author{S. Staeck}
\affiliation{Department of Physics, Technical University of Denmark, Kgs. Lyngby 2800, Denmark.}
\affiliation{European Synchrotron Radiation Facility, 71 av des Martyrs, 383043 Grenoble, France.}

\email{stest@dtu.dk.}

\author{C. Yildirim}
\affiliation{European Synchrotron Radiation Facility, 71 av des Martyrs, 383043 Grenoble, France.}
\author{R. Rodriguez-Lamas}%
\affiliation{ CEA Grenoble/ IRIG/ DEPHY/MEM/NRX, 17 av des Martyrs, 38054 Grenoble Cedex, France.
}%

\author{T. Dufrane}
\affiliation{European Synchrotron Radiation Facility, 71 av des Martyrs, 383043 Grenoble, France.}

\author{C. Detlefs}
\affiliation{European Synchrotron Radiation Facility, 71 av des Martyrs, 383043 Grenoble, France.}

\author{N. Gellert}
\author{A. Gayoso Padula}
\author{H. F. Poulsen}
\affiliation{Department of Physics, Technical University of Denmark, Kgs. Lyngby 2800, Denmark.}
\date{\today}

\begin{abstract}
We introduce the use of a crossed pair of Multilayer Laue Lenses (MLLs) as an objective in Dark-Field X-ray Microscopy (DFXM). In a demonstration experiment at the ID03 beamline at ESRF, two flat Mo-Si MLLs were used, with a physical aperture of $50 \times \SI{50}{\micro\metre^2}$, and a focal length of \SI{14.25}{mm} at \SI{19.0}{ keV}. Applying a 10 \% criterion to the Modulation Transfer Functions (MTFs) acquired, a spatial resolution of \SI{56}{\nano\metre} is obtained in bright-field mode --- more than three times better than with a compound refractive lens (CRL) objective. The dark-field resolution is similar. With an efficiency of 26.7 \% the MLL objective expands the science domain of DFXM significantly, both for bulk and near-surface studies. Similar to the CRL case, the reciprocal space resolution is dominated by the numerical aperture (NA) of the objective, with the NA being three times larger in the MLL case. This enables faster orientation mapping and implies improved options for the use of tomographic reconstruction algorithms. Although the MLL objective pupil varies with energy and position, secondary peaks are suppressed, simplifying both interpretation and forward simulations. We present an example DFXM application using the MLL as objective, imaging a through-silicon via Kelvin device.

\end{abstract}

\maketitle

Hard X-ray imaging is a key tool for non-destructive characterization of crystalline microstructures in bulk materials, providing access to grains, domains, strain fields, and defect structures deep inside engineering alloys, ceramics, rocks and functional devices \cite{Poulsen2001, Ludwig2009, Ice2011, McDonald2015, Poulsen2020, Henningsson2025}. At photon energies above \SI{15}{keV}, diffraction-based contrast enables sensitivity to crystallographic orientation and elastic strain while retaining sufficient penetration depth for three-dimensional studies of embedded micro-structural features. Achieving high spatial resolution under these conditions is essential for resolving fine micro-structural length scales that ultimately govern macroscopic material properties.

Dark-field X-ray microscopy (DFXM) is a full-field diffraction-based microscopy technique that addresses this need by forming a magnified real-space image from a selected Bragg-diffracted beam using an X-ray objective \cite{Simons2015,yildirim2020probing,Poulsen2021}. By varying the scattering vector, maps of local orientation and axial elastic strain can be obtained for a given Bragg reflection. The achievable spatial resolution in DFXM is ultimately determined by the X-ray objective lens. DFXM studies have so far employed compound refractive lenses (CRLs), fabricated from Be \cite{Snigirev1996, Kutsal2019, Lengeler1999, Schroer2005b, Poulsen2017}, diamond \cite{Katze2026} or SU-8 photo-resist \cite{Nazmov2004, Hlusko2020}. A spatial resolution of approximately \SI{150}{\nano\metre} is typically achieved. Although the diffraction limit of Be CRLs can be significantly lower, under realistic experimental conditions lens figure errors and signal-to-noise considerations dominate the attainable resolution.

Multilayer Laue lenses (MLLs) provide a route to overcoming this resolution limit. By exploiting Bragg diffraction in depth-graded multilayer structures, MLLs exhibit substantially higher numerical apertures than refractive optics while maintaining high diffraction efficiency and compatibility with photon energies well above \SI{15}{keV} \cite{Maser2004, Yan2007, Bajt2017, Kubec2018}. State-of-the-art MLLs have demonstrated \SI{3}{\nano\metre} focusing in dedicated nanofocusing configurations \cite{Dresselhaus2024}. MLLs can also be employed as objectives for full-field imaging, as demonstrated in bright-field mode for synchrotron based research \cite{Niese2014, Niese2016, Murray2019} and more recently also for use with laboratory based X-ray sources \cite{Lechowski2024}. While a first demonstration of MLL operation in a dark-field geometry has been reported for a single BaTiO$_{3}$ domain configuration \cite{Kutsal2019}, no assessment of spatial resolution, contrast formation, comparison with CRL-based DFXM image contrast, or broader applicability has been carried out to date.

In this letter, we address this gap by presenting a systematic study of a crossed pair of MLLs for DFXM use. With a fixed mount and a distance of \SI{50}{\micro\metre} between the vertical and horizontal MLL, the pair effectively acts as a single 2D objective. 
We benchmark the intrinsic imaging performance of this objective using bright-field imaging and directly compare DFXM images acquired with MLL and CRL objectives under comparable conditions. The results demonstrate a clear improvement in DFXM spatial resolution beyond the practical limits of CRL objectives.

\begin{figure}
    \begin{center}
    \includegraphics[width=0.95\linewidth]{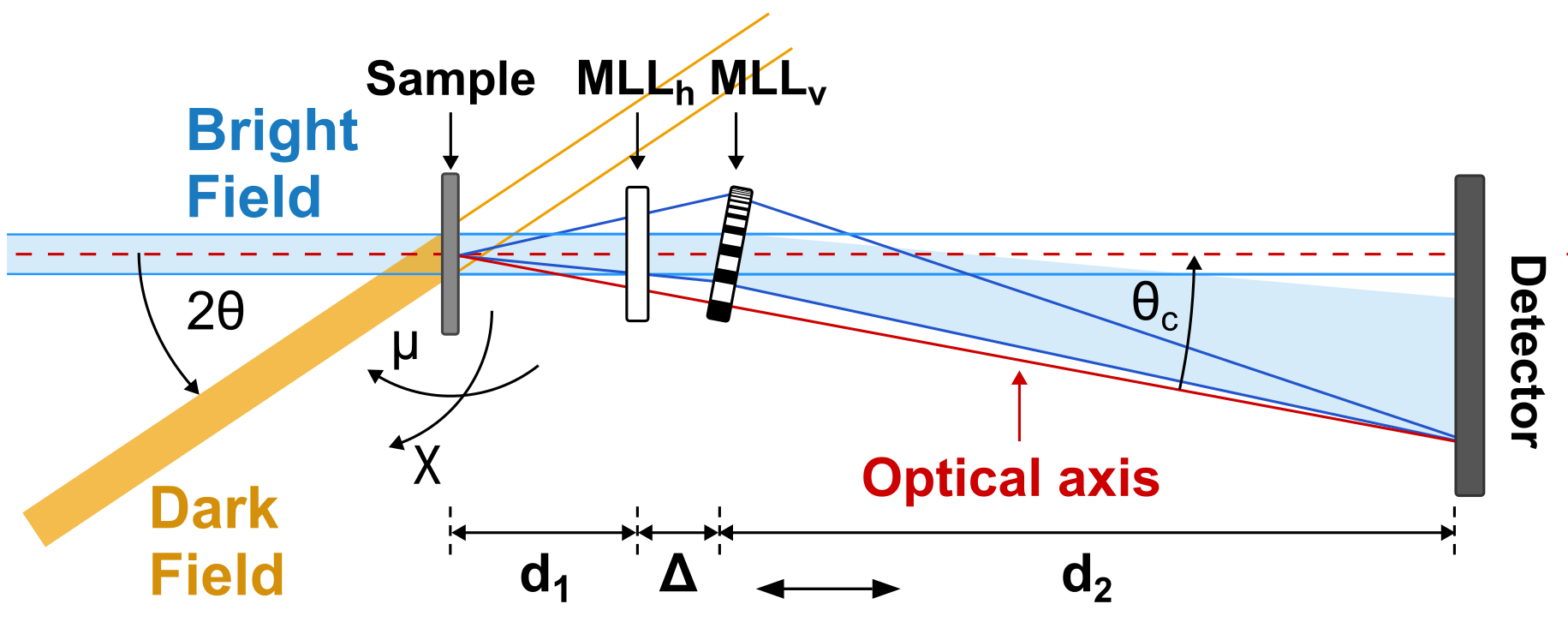}
    \caption{Imaging set-up. MLL$_\mathrm{v}$ and MLL$_\mathrm{h}$ enables imaging by Bragg diffraction in the vertical and horizontal directions, respectively. With a small gap between them, $\Delta = \SI{50}{\micro \metre}$, the two act as one 2D lens objective.  This may be exploited in bright-field (blue) or dark-field (orange) imaging mode. $d_1$ and $d_2$ symbolize the sample-objective and objective-detector distance, respectively. $2\theta$ is the scattering angle related to the diffracting element within the sample. $\theta_c$ is the angle between the incoming beam and the optical axis of the objective. $\mu$ describes the rocking direction motor, $\chi$ the rolling direction motor. Adapted from Murray \cite{Murray2019}.}
    \label{fig-set-up}
    \end{center}
\end{figure}

The applied imaging system is illustrated in Fig.~\ref{fig-set-up} in both bright-field and dark-field mode. The two identical 1D MLLs are oriented orthogonal to each other and used for imaging in the vertical and horizontal directions. For medium to large X-ray magnification $\mathcal{M}$ the angular deviation between the incident beam and the first order diffracted beam, $\theta_c$, is sufficient that no order-defining aperture is needed. 
When $\Delta \ll d_1$ (with distances defined in Fig.~\ref{fig-set-up}), the two MLLs have a joint focal length, $f$.  In this case, $\mathcal{M} = d_2/d_1$, which, combined with the lens makers' equation, leads to the relation $d_1 = (\mathcal{M}+1)/\mathcal{M} \; f$.

The MLL objective's angular acceptance and aberrations preclude a thin-lens description. Following Yan \emph{et al.} \cite{Yan2007} volume diffraction effects can only be neglected if $\omega < (2 \Delta r_{min})^2/\lambda$. Here $\omega$ symbolizes the thickness of the MLL along the X-ray beam direction and $\Delta r_{min}$ the width of the outermost zone. This inequality is not fulfilled for MLL designs of relevance to DFXM.  Hence, the complex pupil (the apparent size of the lens aperture) for the first order diffraction is not fixed but varies with energy, across the aperture and as a function of the incidence angle of the beam in relation to the optical axis of the MLL. To simulate the microscope properties --- and eventually DFXM images --- one must as function of position calculate the fraction of the lens that satisfies the Bragg condition and take into account dynamical diffraction effects.  
Yan \emph{et al.} provided a comprehensive theoretical description using a Takagi-Taupin description for flat and wedged MLLs. Several groups have shown that one may use ray tracing for forward simulations\cite{Kubec2017}. In particular, Chapman and Bajt  used such simulations in conjunction with perturbation analysis to derive off-axis and chromatic aberrations in a focusing geometry \cite{Chapman2020}.\\

%\section{\label{sec:set-up}set-up}

The two identical flat MLLs were manufactured by sputtering subsequent layers of Mo, C, Si and C \cite{Kubec2017}. The detailed layout is provided in Supplementary material, section 1. The two MLLs are oriented orthogonal to each other with a fixed distance of $\Delta = \SI{50}{\micro\metre}$. The combined unit has a physical aperture of $50 \times \SI{50}{\micro\metre^2}$ and a size of $28 \times 20 \times \SI{16}{\milli\metre^3}$. Once assembled, there are no internal degrees of freedom. Fig.~\ref{fig-set-up} shows the basic set-up geometry.

The experiment was carried out at beamline ID03 \cite{Isern2025} at the European Synchrotron Radiation Facility (ESRF). A Si(111) double monochromator was used to select a beam with a photon energy of \SI{19.0}{keV}. The beam was prefocused by a transfocator and shaped with slits, generating a quasi-parallel incident box beam of $\SI{100}{\micro\metre} \times \SI{100}{\micro\metre}$. A decoherer (spinning disk) was placed upstream of the sample to avoid speckle patterns in the images. 
At this energy, the focal length of the objective is $f = \SI{14.25}{\milli\metre}$, and the offset in bright-field between incident beam and the optical axis of the MLL $\theta_c = \ang{0.22}$. With a sample-detector distance of $d_1 + d_2 = \SI{2.5}{\metre}$, the resulting X-ray magnification was $\mathcal{M} = 175$. The X-ray detector was a pco.edge 4.2 far-field camera equipped with  a 10$\times$ objective, corresponding to an effective pixel size of  \SI{3.5}{\nano\metre}. With a reduced ring current of \SI{40}{\milli\ampere}, the exposure time for one image was typically \SI{5}{s}. In addition, a Basler CCD camera with \SI{50}{\micro\metre} pixel size was used --- placed at \SI{2.5}{\metre} from the sample --- in cases where the field-of-view of the pco detector was limiting.

We compare the optical performance of this set-up to that of an CRL objective set-up \cite{Snigirev1996,Roth2014}, at an energy of \SI{17}{keV} and with the same detector. The detailed design of this objective is provided in the Supplementary material, section 2. At \SI{17}{keV} this lens has a focal length of \SI{277}{\milli\metre}, and exhibits a Gaussian angular acceptance with a FWHM of \SI{1.3}{mrad}. The X-ray magnification is $\mathcal{M} = 17.6 $ and the effective pixel size  \SI{37}{\nano\metre}. The ring current was in this case \SI{200}{\milli \ampere} and the exposure time \SI{0.05}{s}.\\

\begin{figure*}
    \begin{center}
    \includegraphics[width=0.7\linewidth]{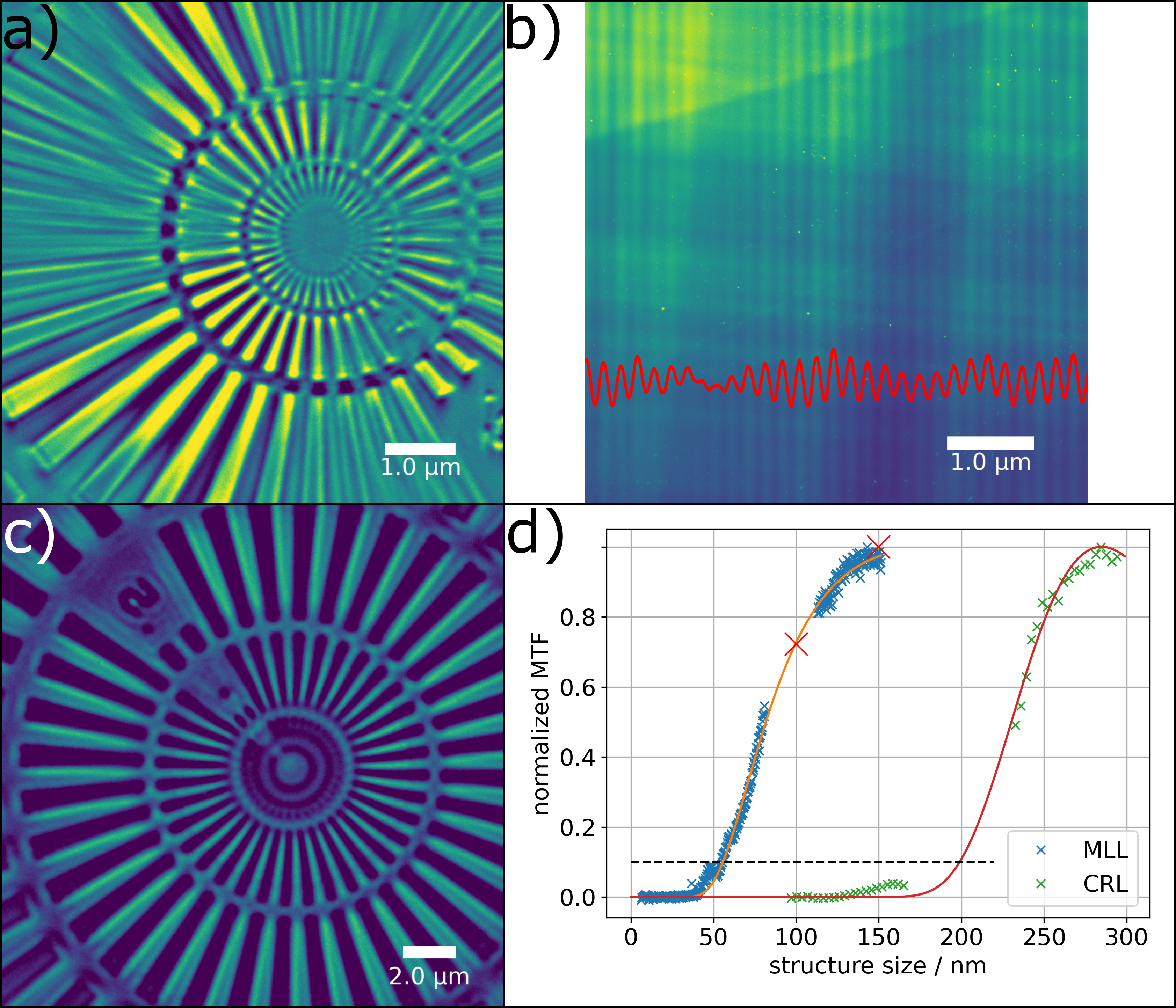}
    \caption{Characterization of the spatial resolution of two objectives.  a) \emph{MLL in bright-field:} Flat field corrected Siemens star recorded with 1852$\times$ total magnification. The innermost structure has a structure size of \SI{50}{\nano \metre}. b) \emph{MLL in dark-field:} Cropped and rotated image of a \SI{100}{\nano \metre} Si single-crystal pattern recorded with a total magnification of 2075$\times$. The red curve represents the 1D modulation of the intensity across the line pattern, summed up over the vertical direction of the image. c) \emph{CRL in bright-field:} Cropped image of the flat field corrected Siemens star recorded with 175$\times$ total magnification. d) \emph{Comparison in bright-field:} Normalized modulation transfer function (MTF) for the wo objectives plotted over the structure size (half-period of the Siemens star). Both MTFs have been fitted with a Gaussian function (red curves) to determine the structure size at 10 \% contrast. The 10 \% value for the MLL is \SI{55.6 \pm 0.3}{\nano\metre}, for the CRL it is \SI{199 \pm 4}{\nano\metre}. 10 \% contrast is marked by the dashed line. The big red crosses mark the contrast in dark-field achieved with the MLL, assuming 100 \% contrast at \SI{150}{\nano\metre}.}
    \label{siemens}
    \end{center}
\end{figure*}

To characterize the resolving power in \emph{bright-field mode}, a Siemens star comprising  a \SI{500}{\nano\metre} thick tantalum 2D pattern was used (NTT AT XRESO-50HC \cite{nttat2026}). The bright-field results for the MLL and CRL set-ups are shown in Fig.~\ref{siemens} a) and c), respectively. A comparison between the two corresponding Modulation Transfer Functions (MTFs) --- shown in Fig.~\ref{siemens} d) --- reveal that the resolution of the MLL objective is more than a factor of 3 better. Using a 10 \% threshold criterion on the MTF, the resulting resolution is \SI{56}{\nano\metre} for the MLL objective.  (The 10 \% criterion is less applicable to the data set of the CRL). For both set-ups the response is to a first approximation isotropic. 

For an estimation of the resolving power in \emph{dark-field mode}, we chose a set of 1D test patters, manufactured on a Si wafer by the CEA \cite{Borel2023, Nicolas2024}. The structures are cylindrical holes in the surface filled with copper, organized to make line patterns. The lateral size of these holes varies between \SIrange[]{100}{500}{\nano\metre} for different areas on the sample wafer, while their depth is \SI{6}{\micro\metre}. The distance between the holes is equal to their lateral size, so that a \SI{100}{\nano\metre} hole pattern features a \SI{200}{\nano\metre} period in real space.

For dark-field mode, the sample was imaged in reflection geometry, using the Si 400 reflection. The resulting DFXM image for the \SI{100}{\nano\metre} pattern is shown in Fig.~\ref{siemens} b). The linear structures are clearly visible. The contrast at \SI{150}{\nano\metre} is slightly better, but it does not improve any further when going to \SI{400} and \SI{500}{\nano\metre}. The ratio in DFXM contrast between \SI{150}{\nano\metre} and \SI{100}{\nano\metre} is identical to that in bright-field, see Fig.~\ref{siemens} d). Hence, we argue that the spatial resolution in dark- and bright-field is the same in the direction orthogonal to the beam (left-right in Fig.~\ref{siemens} b).

The properties of the MLLs degrade toward the edges, displaying interference-like fringes. 
The effective \emph{Field-of-View}, FOV, was determined in bright-field to be \SI{44.2 \pm 0.5}{\micro\metre} in the horizontal direction, in the vertical direction it is \SI{43 \pm 3}{\micro\metre}. The combined efficiency of the two MLLs was 26.7 \% , corresponding to simulations made by the manufacturer.
In order to characterize the image quality and resolution over the whole FOV of the MLL objective, we determined the spatial resolving power as a function of position in the sample. As discussed in the Supplementary material, section 3, within experimental error the response is uniform.

The MLLs are associated with linear structures, that are visible as dark lines in the flat field. These are analysed in the Supplementary material, section 4, where it is shown that they effectively can be removed by normalization.\\

The reciprocal space resolution function is a key feature of any diffraction-based imaging instrument, as it determines properties such as local strain and local orientation resolution, FOV and options for mapping defect densities. For DFXM based on a CRL objective, these properties can be assessed with excellent accuracy by geometrical optics \cite{Poulsen2017,Poulsen2021}, which have been validated in both monochromatic and pink beam data acquisition modes \cite{Yildirim2025, Labella2025}. 
In essence, on-axis the reciprocal space resolution function has the shape of a disc, with a shape that in two directions is defined by the Gaussian angular acceptance function of the CRL, with a width that equals the NA. Off-axis the function simply shifts in reciprocal space \cite{Poulsen2021}.

To determine the reciprocal space resolution function in the MLL case, a Si single crystal was used as a test specimen, diffracting on the (220) reflection with $2\theta = \ang{19.55}$. For this experiment, the Basler far-field detector was used, to provide full integration. Following standard DFXM procedures, three types of scans were performed \cite{Poulsen2021}, see Fig.~\ref{fig-set-up}: 

\begin{itemize}
\item \emph{Rocking scan.} Scanning the base-tilt, the $\mu$-motor, corresponding to varying $\theta$ in a classical $\theta-2\theta$ set-up. 
\item \emph{Rolling scan.} Scanning the $\chi$-motor, tilting the diffraction vector around the incident beam direction. 
\item \emph{Longitudinal scan} A 2D scan, where in the outermost loop  the pitch of the MLL objective is varied, corresponding to varying the scattering angle $2\theta$. Within this loop a rocking scan is performed. For each pixel on the detector the intensity is averaged over $\mu$. 
\end{itemize}

Suitably normalized, these scans characterize the angular resolution along three nearly orthogonal strain directions \cite{Poulsen2021}:
$q_{\mathrm{rock}}', q_{\mathrm{roll}}$ and $q_{\perp}$.

\begin{figure*}%
    \begin{center}
    \includegraphics[width=1\linewidth]{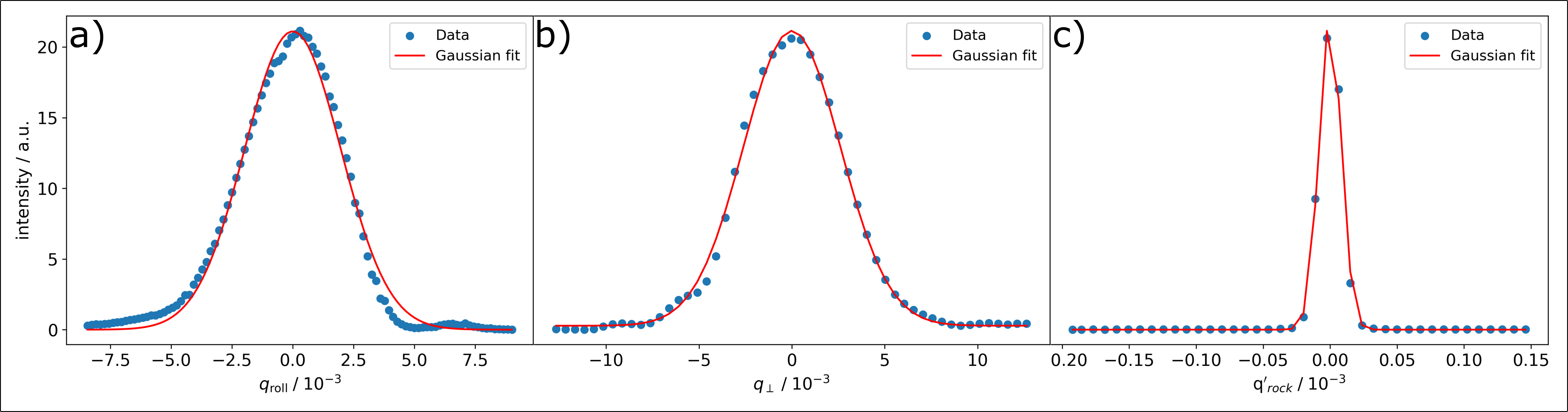}
    \caption{Components of the reciprocal space resolution function of a Si (220) reflection at \SI{19}{\kilo\electronvolt}, shown in dimensionless (strain) units.  a) The rolling direction, $q_{\mathrm{\mathrm{roll}}}$, b) the longitudinal direction, $q_{\perp}$ and c) the rocking direction,  $q'_{\mathrm{rock}}$. Shown in red are best fits to simple Gaussian models.}
    \label{rec_space}
    \end{center}
\end{figure*}

The results averaged over the entire image on the detector are shown in Fig.~\ref{rec_space}. In the longitudinal direction, Fig.~\ref{rec_space} b),  the resolution function appears to be reasonably well approximated by a Gaussian. The FWHM is $\Delta q_{\perp} = \frac{\Delta Q_{\perp}}{Q_0} = \SI{0.0062 \pm 0.0001}{}$. This is in stark contrast to the behaviour of a simple zone plate,  which would give rise to a top-hat function. The apparent smearing is due to different subvolumes of the MLL being in the diffraction condition at different tilt angles. As expected due to symmetry, the results for the rolling direction, Fig.~\ref{rec_space} a), are similar. The FWHM is in this case  $\Delta q_{\mathrm{roll}} = \SI{0.0055 \pm 0.0001}{}$.

Notably, both FWHMs are approximately three times larger than the corresponding FHWM for the CRL set-up introduced above \cite{Labella2025}: $\Delta q_{\perp} = \Delta q_{\mathrm{roll}}  \approx \SI{0.0020}{}$. For the rocking scan, a FWHM of $\Delta q'_{\mathrm{rock}} = \SI{1.9 \pm 0.3}{} \times 10^{-5}$ is similar to that for the CRL based set-up, indicating that the limiting factor in this direction is not the objective but the Darwin width of the sample.

An analysis of the variation of the reciprocal space resolution with position in the sample plane is provided in the Supplementary material, section 4. It shows that the diffraction condition of the MLL is never fulfilled everywhere at once in the aperture and that the pupil of the MLL is not constant, but varies with position in the aperture. However, the resolution functions do not display side lobes as in bright-field measurements from previous work \cite{Kubec2017, Murray2019}. It is a major finding of this work that such side lobes are much suppressed in the current set-up, vastly simplifying the interpretation of data.\\

As a use case, a through-silicon via (TSV) Kelvin device \cite{Borel2024} manufactured by CEA was imaged with DFXM using the MLL objective and compared to a prior measurement of the same device with the CRL objective. TSVs are used to connect stacked semiconductor chips, the Kelvin device is a structure manufactured on a TSV wafer to measure the resistance of a single TSV for reference. Here, a region of \SI{1}{\micro\metre} wide, \SI{6}{\micro\metre} thick parallel electrical connections was imaged.

In both cases, the sample was imaged in reflection geometry, using the Si (400) reflection at 19.0 keV.  
In the CRL case, the effective pixel size was \SI{51}{\nano\metre} and the total magnification 128. In the MLL case, a 2$\times$ optical objective was used, to better match the pixel size of the CRL set-up: it was \SI{15.7}{\nano\metre}.

\begin{figure}%
    \begin{center}
    \includegraphics[width=0.95\linewidth]{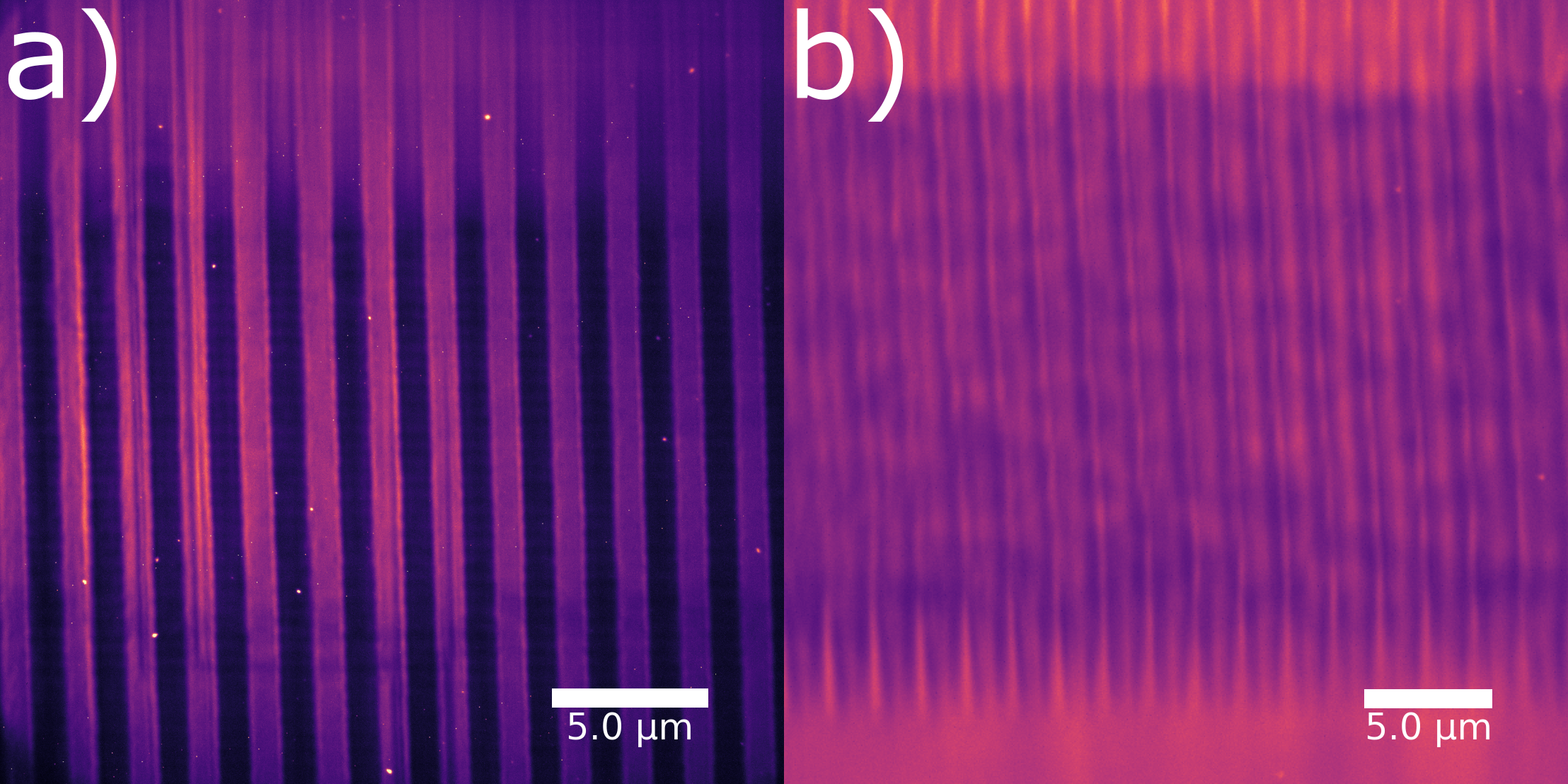}
    \caption{KEL2 device imaged in DFXM reflection geometry with a) the MLL and b) the CRL at ID03. The structure shown is some parallel connectors, running from top to bottom. The width of these structures is \SI{1}{\micro\metre}. The direction of the beam is in both cases parallel to up-down in the images. Both images are cropped for presentation.}
    \label{KEL}
    \end{center}
\end{figure}

Images representing approximately the same region-of-interest are compared in Fig.~\ref{KEL}. 
Resolution and contrast are clearly superior in the MLL image.\\

Improving spatial resolution is of interest for any microscope in materials science, as most structures are inherently multiscale. Significant progress has been demonstrated in relation to improving X-ray microscopy using CRLs. In particular Wang \emph{et al.} recently demonstrated integrated aberration compensation in 
diamond X-ray lenses, thereby generating focused beams of \SI{50}{nm} size \cite{Wang2025}. Still, CRL manufacturing  makes this a difficult route to progress on resolution.  

In comparison the spatial resolution is easier to improve using MLL optics. By decreasing the focal length, the spatial resolution can be improved --- so far \SI{3}{nm} have been demonstrated in bright-field with wedged MLLs \cite{Dresselhaus2024}. Furthermore, phase correction plates can be employed to improve spot sizes \cite{Seiboth2022}. Our work demonstrates that the resolution  in dark-field with an MLL objective matches that of bright-field and hence suggests a way to tailor the spatial resolution also in dark-field. In addition, the objective used is compact, and may as such be added as a complementary optics tool for multiscale studies. 
The MLL objective provides  several additional \emph{advantages} with respect to existing CRL objectives:  

\begin{itemize}

 \item \emph{Faster orientation mapping.} For samples with an orientation spread significantly larger than the objective NA, such as deformed metals\cite{Mavrikakis2019, yildirim20224d}, the angular sampling provided by DFXM is often too fine. While the step-size for scanning in the rocking direction can be adjusted at will by continuous scanning, the step-size in the rolling direction is constrained by the NA. Hence, all else being equal, for mosaic samples the MLL may reduce the number of angular steps. 
 %by roughly a factor of three. 
 Likewise the larger NA in the $2\theta$ direction implies that one effectively integrates over the entire relevant $2\theta$ range, whereas the smaller NA for the CRL case implies that one may need to superpose several images, acquired at different $2\theta$'s, even if it is only the orientation map that is of interest. 

  \item \emph{Improved conservation of integrated intensity in angular scans.}  Although DFXM was originally intended to operate in box-beam mode and provide 3D information using tomographic reconstruction routines \cite{Simons2015}, it has been difficult to sample projections in a robust way. Notably, such topo-tomography routines \cite{Simons2015, shukla2025} require that the integrated intensity is conserved. By inserting a square aperture in the Back Focal Plane (BFP) of the objective, the longitudinal and rolling direction resolution functions can be made approximately top-hat functions. Stitching in angular space is much easier when the NA is a top-hat function.

   \item \emph{Support for coherent diffraction imaging with an objective lens.}  As spatial resolution improves, the voxel size approaches the coherence length --- a trend amplified by increasing source brilliance. Pedersen \emph{et al.} demonstrated how DFXM can be generalized to use a coherent beam \cite{Pedersen2020a};  iterative oversampling routines and Fourier synthesis were used to reconstruct shape and strain from the far-field intensity pattern. 
   
   The MLL's ability to preserve the phase front \cite{Morgan2015} is here relevant.

    \item \emph{Enhanced energy range.}  The efficiency of the MLLs is generally speaking better than CRLs at energies above \SI{30}{keV}. This is in part due to the fact that refractive and diffractive lenses have focal lengths that increase with the X-ray energy $E$ as  $E^2$ and $E$, respectively. 

     \item \emph{Magnification.}  The much shorter focal lengths imply that an MLL based set-up will involve an order of magnitude increase in X-ray magnification. 
     This enables the use of more efficient photon counting detectors.
\end{itemize}

The main \emph{disadvantage} of the use of MLLs is the trade-off between spatial resolution and working distance. The short working distance limits the range of sample environments available for convenient studies of bulk materials. Moreover, for dose-sensitive materials, it is problematic that most of the diffracted signal does not reach the detector. 

Another disadvantage is the deterioration of the angular resolution in the rolling and longitudinal directions, cf.~Fig. \ref{rec_space}. This will \emph{disadvantage mapping of strain components} in the two directions and makes it impossible to derive defect densities. The remedy may again be the insertion of an aperture in the Back Focal Plane of the objective, similar to what is done with CRL objectives \cite{Staeck2026} --- to be explored elsewhere. 

Finally, we argue that a spatially dependent reciprocal space function  is significantly harder to model. This will complicate attempts to perform forward modeling.\\

This work demonstrates the use of two cross-linked MLLs as a single objective for DFXM use. MLLs are intrinsically not limited by manufacturing errors in the same way as CRLs, and hence this work opens up for DFXM studies at finer length scales, limited ultimately by signal-to-noise considerations and working distance. Moreover, the MLL optics used is a compact device, that is well suited as part of a multiscale microscope, encompassing grain mapping (3DXRD), domain mapping (DFXM with CRLs) as well as a defect mapping modalites (DFXM with MLL).  

Another distinctive difference is the much increased numerical aperture of the MLL set-up - which again scales with focal length. This is an advantage for mapping of microstructures exhibiting characteristic orientation changes that are larger than the NA, but a disadvantage for strain and in particular defect density studies as these rely on high angular resolution.\\

We thank Sven Niese for scientific discussion and ESRF for providing the beamtime at ID03. We thank Merve Kabukcuoglu for providing the sample for the reciprocal space resolution measurements. We thank Marylin Allegra for dicing the 1D dark-field test sample and the TSV Kelvin device, Rémi Quenouillère for the lithography process, Antonio Roman for the etching and Michaël Bouvier for the deposition of those samples. We acknowledge financial support from the ERC Advanced Grant nr 885022, from the Danish ESS lighthouse on hard materials in 3D, SOLID as well as from the Villum Investigator grant  nr.~73771. Can Yildirim acknowledges support from the ERC Starting Grant nr 10116911.

\section*{Author Declarations}

\subsection*{Conflict of Interest}
The authors have no conflicts to disclose.

\subsection*{Author Contributions}
\textbf{S. Staeck:} Data Curation; Formal Analysis; Investigation; Software; Visualization; Writing -- original draft. \textbf{C. Yildirim:} Conceptualization; Funding acquisition; Investigation; Methodology; Project Administration; Resources; Supervision; Visualization; Writing - review \& editing. \textbf{R. Rodriguez-Lamas:} Conceptualization; Investigation; Resources; Validation; Writing -- review \& editing. \textbf{T. Dufrane:} Resources. \textbf{C. Detlefs:} Conceptualization; Funding acquisition; Investigation; Methodology; Project Administration; Resources; Supervision; Writing -- review \& editing. \textbf{N. Gellert:} Investigation. \textbf{A. Gayoso Padula:} Investigation. \textbf{H. F. Poulsen:}
Conceptualization; Funding acquisition; Investigation; Methodology; Project Administration; Resources; Supervision; Visualization; Writing -- original draft; Writing -- review \& editing.

\section*{Data availability}
Data presented in this paper can be obtained upon reasonable request.

\section*{REFERENCES}

%\nocite{*}
\bibliography{mll}% Produces the bibliography via BibTeX.

\clearpage
\appendix
\section*{Supplementary Material}
\subsection{MLL details}
The two identical flat MLLs, which are part of the 2D MLL presented in the main document, were manufactured by sputtering subsequent layers of Mo, C, Si and C \cite{Kubec2017}. The C layers act as buffers with a fixed thickness of \SI{1}{nm}. The thicknesses of the Mo and Si layers are equal but vary with distance from the optical axis, such that the total Mo-C-Si-C layer has a thickness compatible with the period of a zone plate. The periods correspond to zones 971 to 6970,  resulting in a total thickness of $D = \SI{50}{\micro\metre}$, smallest and largest period thicknesses of \SI{11.55}{\nano\metre} and \SI{30.95}{\nano\metre}, respectively, and outermost and innermost zone widths of $\Delta r_n =  \SI{5.78}{\nano\metre}$ and \SI{15.48}{\nano\metre}. The MLLs were manufactured each with a section thickness (dimension along the beam direction) of $\omega = \SI{9.6}{\micro\metre}$.

\subsection{CRL details}
The CRL objective currently in use at ID03 comprises 87 Be lenses, each with a physical aperture diameter of \SI{440}{\micro\metre}, a radius-of-curvature-at-apex of  $R = \SI{50}{\micro\metre}$, and a web thickness of \SI{30}{\micro\metre}. The distance between the center-of-mass of lenses is \SI{1.69}{\milli\metre} on average.

\subsection{Spatial variation of the direct space resolution}
\begin{figure}
    \begin{center}
    \includegraphics[width=0.85\linewidth]{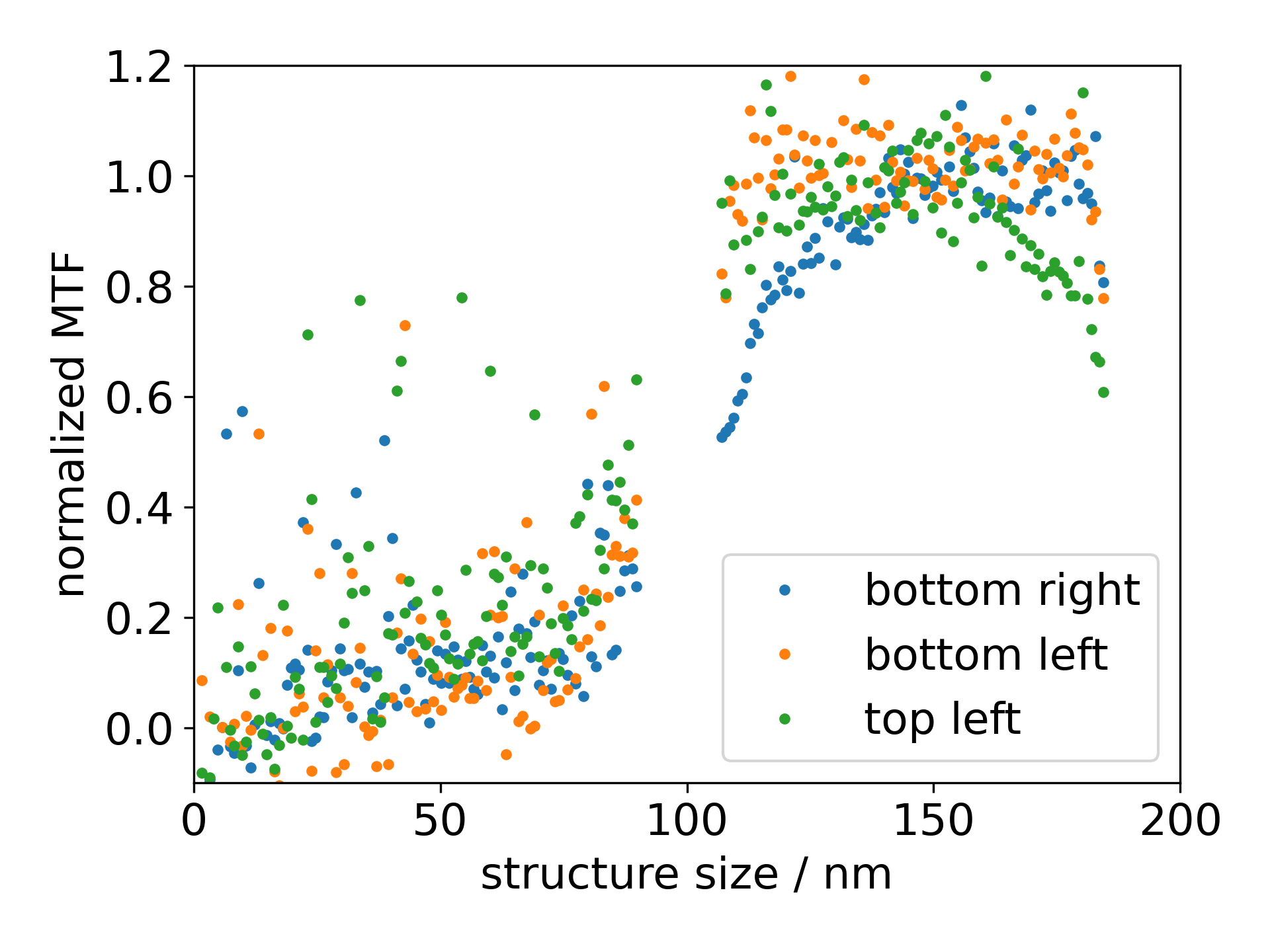}
    \caption{Characterization of spatial resolution for different positions in the MLL's FOV.}
    \label{MTF_FOV}
    \end{center}
\end{figure}

For the evaluation of the heterogeneity of the direct space resolution of the MLL, we repeated the measurement shown in Fig.~3 of the main document. The Siemens star test structure was imaged in direct beam at \SI{19}{keV}, with a \SI{5}{\metre} sample-to-detector distance and the 2x objective of the pco far-field detector. The exposure time was \SI{3}{\second}; 20 images have been acquired at each position and averaged. The images are corrected with 20 dark frames and 20 flat fields. The Siemens star pattern was imaged at three different positions of the FOV of the MLL, designated bottom right, top left and bottom left, referring to the position in the FOV. The results are shown in Fig.~\ref{MTF_FOV}. 

The data is rather noisy, most likely because the signal is spread over very few pixel in the 2x pco far-field detector. Apart from that, the data shows that there are no significant differences within experimental error in terms of resolving power between the different MLL areas. Thus, it can be concluded that a constant imaging quality is maintained over the whole FOV of the MLL. Note that these measurements were performed with a different objective mounting, that was more prone to vibrations. Probably because of that, the resolution that was achieved in this measurement is lacking compared to that presented in Fig.~2 of the main document.  

\subsection{Flat field correction algorithm}

\begin{figure}
    \begin{center}
    \includegraphics[width=0.95\linewidth]{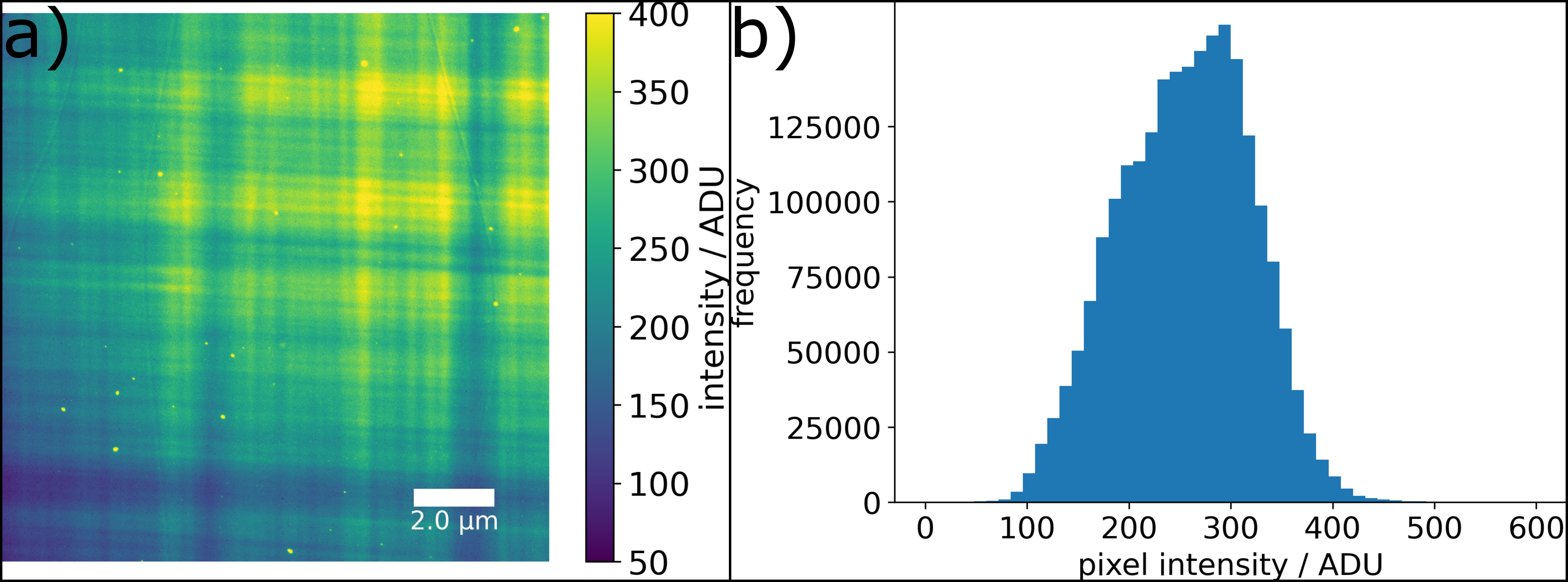}
    \caption{a) Flat field within a region of interest of $13.8 \times \SI{13.8}{\micro\metre\squared}$ for an MLL objective with the same specifications as used in the rest of this work, recorded at ID03, ESRF. A dark frame was subtracted beforehand. b) Histogram of the intensity distribution in a). The axis has been limited to 600 analog-to-digital units (ADU) to cut off hot pixel.}
    \label{flat field}
    \end{center}
\end{figure}

A characteristic bright field image generated by the MLL objective in response to a nearly homogeneous incident beam is shown in Fig.~\ref{flat field}  a). The line structures are an intrinsic effect of the manufacturing. 

\begin{figure*}
    \begin{center}
    \includegraphics[width=0.65\linewidth]{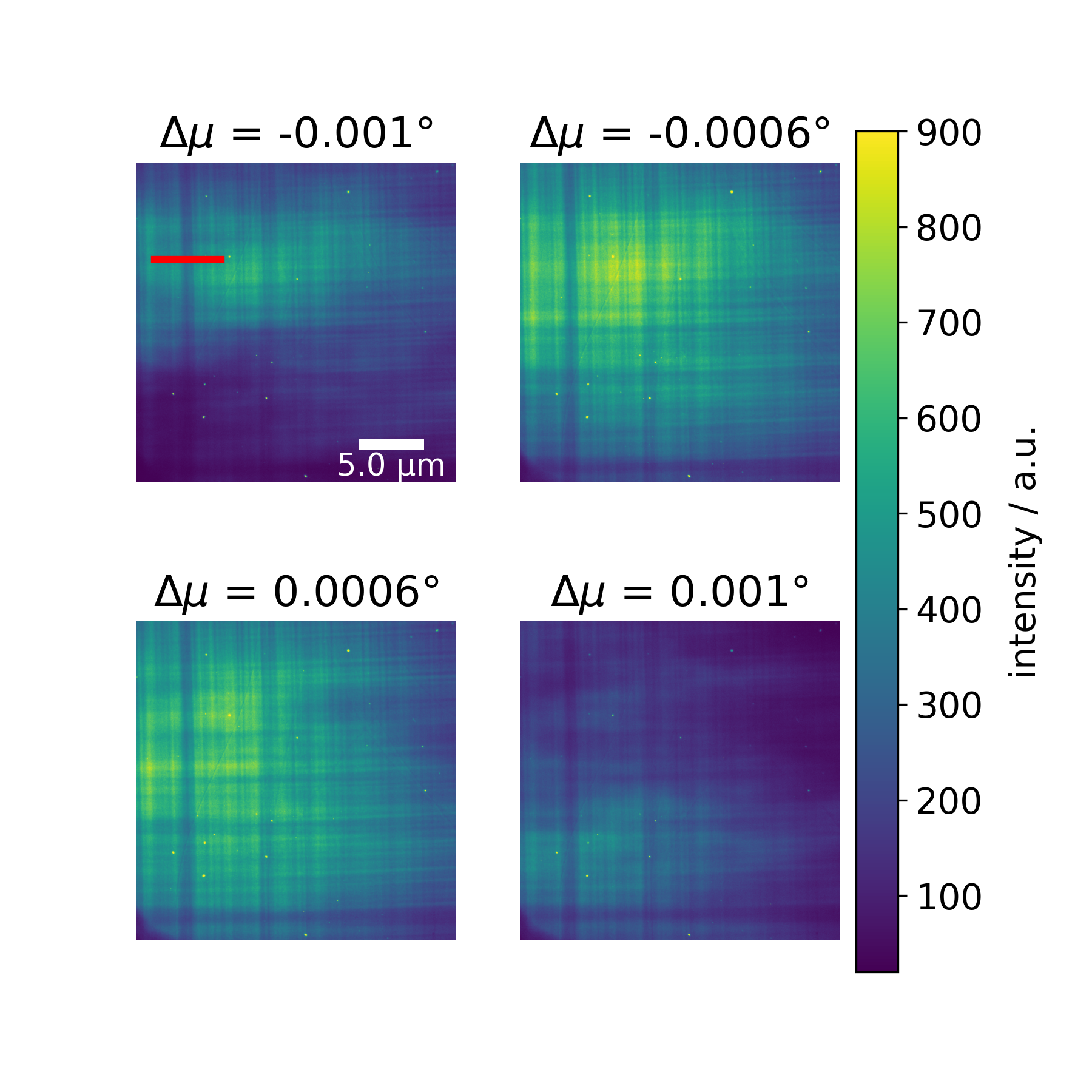}
    \caption{Four different images acquired during a rocking scan recorded in dark-field mode with a Si wafer as the sample. The position on the rocking curve,  $\Delta \mu$, is defined with respect to the position where the integrated intensity over the entire image is at its maximum. The image section used to evaluate the intensity changes around a single dark line is marked in red in the first image. The images are all corrected by subtracting the average of 20 frames acquired with no beam.}
    \label{flatfield_plot}
    \end{center}
\end{figure*}

While the variation of the efficiency in a given region of the objective overall depends on $\theta_c$ (or the angle of the incoming beam, expressed in terms of the rocking direction motor $\mu$), the relative intensities around a given dark line are found to be constant with a standard deviation between 1 \% and 5 \%. Hence, to calibrate for the dark lines we can use the bright field signal integrated over $\theta_c$, $I_{MLL}$ and from this define a flat field, $I_{ff}$ as

\begin{equation}
I_{ff} = \frac{I_{\mathrm{MLL}} - I_{\mathrm{no\ beam}}}{I_{\mathrm{no\ MLL}} - I_{\mathrm{no\ beam}}}.
\end{equation}

Here I$_{\mathrm{no\ MLL}}$ and I$_{\mathrm{no\ beam}}$ represent similar images acquired without the MLL objective and without any beam, respectively.
This image can then be used as a conventional ``flat field'' image also for DFXM. 

In order to verify the above approach for flat field correction for DFXM images, the dependency of the MLL image artifacts - the dark lines - on the rocking angle is investigated. The diffraction response of a Si wafer can be assumed to be sufficiently homogeneous, so that no visible structures from the sample influence the resulting image. The (220) reflection of this wafer was imaged at \SI{19}{keV} in Fig.~\ref{flatfield_plot}. 

Selected images are displayed in Fig.~\ref{flatfield_plot}. The relative intensity distribution in the recorded images shifts with position as a function of $\Delta \mu$. To test whether the contrast around a vertical or horizontal dark line in the images scales with the local intensity, horizontal line profiles summed over 10 pixel rows are analyzed. The area, from which these line profiles are derived, is shown as a red square in the first image in Fig.~\ref{flatfield_plot}. 

\begin{figure*}
    \begin{center}
    \includegraphics[width=0.7\linewidth]{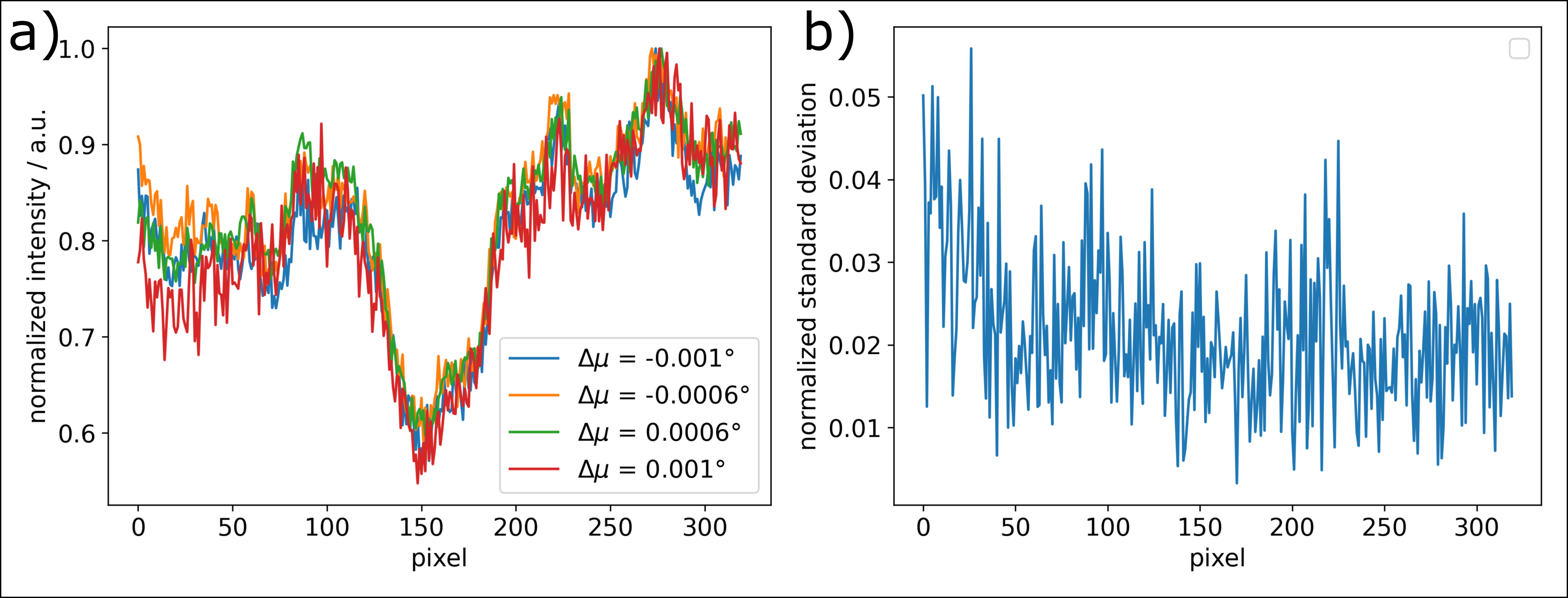}
    \caption{a) Line profiles, derived from the images shown in Fig.~\ref{flatfield_plot},  for different rocking angles $\Delta \mu$, relative to the maximum value of the rocking curve. The intensity has been normalized. b) Standard deviation of the normalized line profiles.}
    \label{flatfield_analysis}
    \end{center}
\end{figure*}

The resulting line profiles are shown in Fig.~\ref{flatfield_analysis} a). When normalized, 
the line profiles are nearly identical. The standard deviation of the line profiles is depicted in Fig.~\ref{flatfield_analysis} b) for a more quantitative analysis. The standard deviation differs between 1 \% and 5 \%, thereby validating the assumption made in the main document, that the relative intensity profile around a dark line is constant.   

\subsection{Reciprocal space resolution analysis}

\begin{figure*}%
    \begin{center}
    \includegraphics[width=0.7\linewidth]{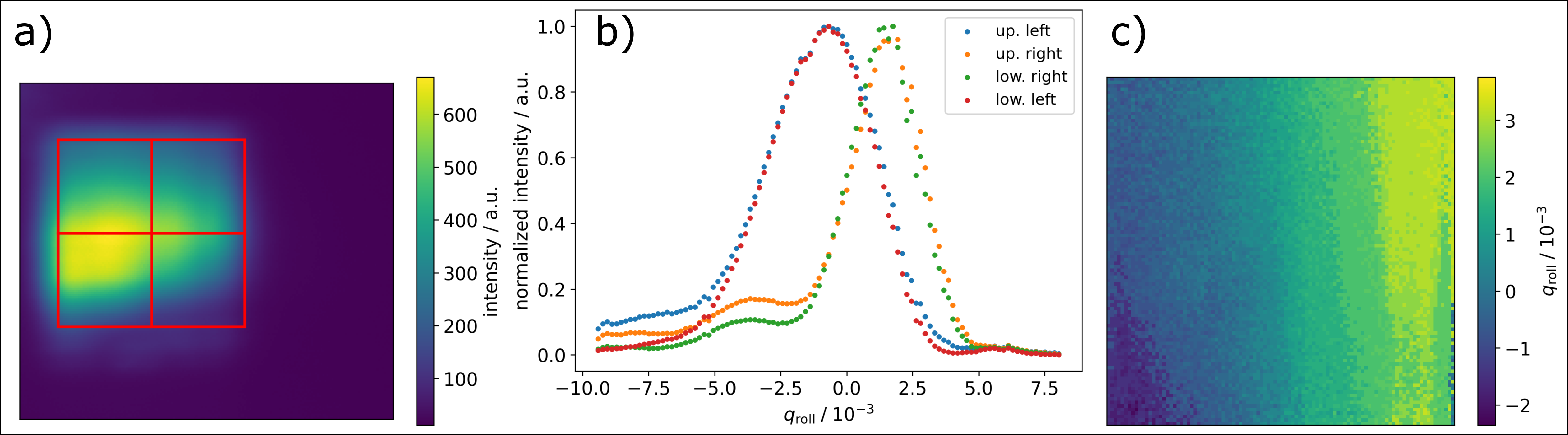}
    \caption{Characterization of the spatial variation of the rolling component of the reciprocal space resolution function. a) Identification of four sub-regions within the entire field-of-view.  b) The $q_{\mathrm{roll}}$ resolution functions determined in the same way as in Fig.~3 a) of the main document for the four quadrants. c) The maximum value of $q_{\mathrm{roll}}$ for each pixel within the entire field-of-view as defined by the outermost box in a).}
    \label{roll_an}
    \end{center}
\end{figure*}

In this section, we characterize the variation of the reciprocal space resolution function with position on the screen. As illustrated in Fig.~\ref{roll_an} a), we divide the region-of-interest into four quadrants and repeat the analysis of the rolling scan for each of these quadrants. The resulting angular acceptance curves are shown in Fig.~\ref{roll_an} b). Furthermore, the $q_{\mathrm{roll}}$ value of the maximum intensity of each pixel within the ROI is plotted in Fig.~\ref{roll_an} c).

We deduce from Fig.~\ref{roll_an} that the strain resolution function $q_{\mathrm{roll}}$ --- as expected --- is independent on the position in the up-down direction. Considering the center position in the $q_{\mathrm{roll}}$ range, this varies in a near linear fashion with left-right position. This gradient reflects that the optical axis moves - the same gradient appears in rolling scans with a CRL based objective. 

The major difference to the CRL based set-up is that the shape of the curve changes when going from left to right, cf. Fig.~\ref{roll_an} b). This reflects that the pupil is not constant but varies as a function of where in the MLL the local diffraction condition is fulfilled.  Previous work has reported on the appearance of side lobes when going off axis in bright field mode \cite{Kubec2017, Murray2019}. These side lobes are not present in the resolution functions shown in this work.\\

\end{document}